\documentclass[12pt]{article}
\usepackage{amsfonts}

\newcommand{\ppp}{\(\psi^{++}\)}
\newcommand{\cpp}{C\raisebox{.2ex}{\small{++}}}
\newcommand{\oReg}{$^{^{\circledR}}$ }

\title{Object oriented and functional programming
       for symbolic manipulation}
\author{Alexander Yu.\ Vlasov  \\
        Federal Center for Radiology, IRH, \\
         Mira Street, 8,
         St.--Petersburg, 197101, Russia
\footnote{E-Mails: alex@protection.spb.su, vlasov@physics.harvard.edu}}
\date{}

\begin{document}

\maketitle

\begin{abstract}
The advantages of mixed approach with using different kind of
programming techniques for symbolic
manipulation are discussed. This type of programming environment now is in
development. The work is stimulated by the necessity of managing
complicated algebraic expressions and formulae of noncommutative
geometry related with HENP.

Functional languages are convenient for representation and
manipulation of symbolic data. From the other hand, it is
convenient to use special ``shell'' around the functional language to
make programming more transparent.

The main purpose of  approach offered is merge the methods of
object oriented programming that convenient for
presentation data and algorithms for user with advantages of functional
languages for data manipulation, internal presentation, and portability
of software.

\end{abstract}


\newpage

\section{Introduction}    

We can consider two different approaches to {\em computer algebra\/}. On the
one hand, it may be application of usual programming techniques to specific
tasks and it is not essential to determine the particular programming language we are
to use. The main problem here is to express the task
by usual simple data types and operations. For example, we can express
monomials like $x_1^k x_2^{l-k}\overline{{y_1}^m}\overline{{y_2}^{n-m}}$ those
form \mbox{$(l/2,n/2)$} representation of Lorentz group by using data
{\em ``register''\/} with four numbers like
\raisebox{-.5ex}{\frame{$\makebox[1.2em][c]{\it k}|\makebox[1.2em][c]{\it l}\|
\makebox[1.2em][c]{\it m}|\makebox[1.2em][c]{\it n}$}} .
Multiplication of the monomials is addition of the {\em registers\/}, complex
conjugation is swapping of left and right parts, etc. . If we work with more
complicated structures, the {\em ``advanced''\/} properties of programming
language can be more useful.

One of such properties is the constructive conception of {\em data types\/}
as some preliminary information about data accessible for translator
and possibility to construct new {\em data types\/} and {\em data
structures\/} in addition to basic types \cite{Wirth76} .

Other properties are discussed further. An object oriented Pascal-like
language with some {\em functional objects\/} is used for the demonstration of
possible synthesis of {\em object oriented\/} and {\em functional\/} programming.
It is called here \ppp\ and is used in examples.


\section{Object Oriented Programming (OOP)}

The OOP languages are successfully used in modern computer science. Let us
consider standard of \cpp\ \cite{Stroustrup86} or some version of
{\em Object Pascal\/} \cite{Delphi95}.

The object oriented style supposes some principles of decomposition of
complicated tasks:

\begin{itemize}
\item The basic structure is {\em an object\/}. The {\em object\/} encloses
 not only {\em data fields\/}, but {\em methods\/} (procedures, functions)
 for manipulation with these data.
\item An object can be defined as {\em descendant\/} of some other object.
 The object {\em inherits\/} all fields and methods of {\em ancestor}.
\item {\em Descendant objects\/} can be used anywhere as parameters of functions
 or right part of {\em assignment\/} operator instead of {\em ancestor}.
\end{itemize}

An analogy of OOP with language of mathematics should be mentioned. It is
useful for applications to {\em computer algebra}.
For example we can define object {\em Group\/} with two methods:

\medskip

{\obeylines
Group = \underline{\bf Object};
\quad \underline{\bf function} \underline{\bf infix} $+$(A,B : Group) : Group;
\{ \cpp\ version:  {\sf friend Group operator + (Group,Group)} \}%
\footnote{Within exampes comments are inclosed in brackets: `\{' and `\}' }
\quad \underline{\bf function} \underline{\bf prefix} $-$(A : Group) : Group;
\underline{\bf end}; \{ Group \}
}

\medskip

It is possible to define $Algebra = \underline{\bf Object}(Group)$
with new multiplicative operation `$*$', {\em Complex\/} numbers, {\em Dirac algebra\/} etc.\ .
Due to properties of OOP it possible to use function `$+$' for any descendant
of {\em object Group\/} and both functions `$*$', `$+$' for any descendant of
{\em Algebra\/}. The methods like `$*$' can be different for all descendant
objects, but expression like $A*B$ invokes necessary algorithm for particular
objects A and B due to standard principles of OOP\@. In the \cpp\ the operator$*$
can be {\em overloaded\/} for any object.

Another useful ``algebraic'' property of languages like \cpp\ is
the possibility of {\em type conversion\/}. It is possible to write
expression like $x:=2*(1+2*i)$ instead of $x:=Complex(2,0)*Complex(1,2)$
because {\em real numbers\/} like $2$ will be converted to
$(Re:2,Im:0)$ of type {\em Complex\/} by constructor Complex($x$ : Number).

\section{Functional Programming (FP)}

The FP languages like {\sc LISP} were one of the first tools for
{\em symbolic manipulation\/} and {\em artificial intelligence\/} (AI).
One of advantages of LISP for such applications is {\em list\/}
data structure. On the other hand {\em lists\/} also can be simple defined
in most of modern imperative languages.

Another advantage of FP is the clear functional structure of
programs \cite{Field88}.  In FP the whole program is considered as composition of
functions. The {\em ``pure''\/} functions are transformations of {\em
argument} to {\em result\/} and do not depend on some global
external variables, unlike the functions and procedures in most of programming
languages.

Due to the pure functional structure in most of FP languages there are very
powerful possibilities of construction and evaluation of {\em new functions\/} at
{\em run time}. It is possible to construct a new function as some special
structure and perform it by using a special operator like {\sf EVAL} in LISP.
This property is very useful for AI due to possibility of {\em ``learning''\/}
by constructing of new {\em algorithms}. In {\em symbolic manipulation\/} the
building of new functions is necessary for
{\em interpretation\/} of symbolic expressions related with some variables and operations.
For example, the string like {\sf abc} often means {\sf OP(a,OP(b,c))} there
{\sf OP} is some function and it is necessary to have the ability to
perform these {\sf OP} operations when values have been assigned to the
variables {\sf a}, {\sf b} or {\sf c}.

\section{Combination of OOP and FP styles}

The FP and OOP languages are mentioned above are {\em universal\/} languages.
Any algorithm that has been written by using one of them can be
rewritten by using any other universal language. The only problem is
complication.

For example, it is possible to define
$Polynomial=\underline{\bf Object}(Algebra)$
and overload operators `$*$' and `$+$' for working with symbolic
expressions for polynomials in OOP languages like \cpp .

There some synthesis OOP and FP is possible because both these approaches
have tendency to unification of data structures and algorithms.
In OOP {\em data fields\/} and {\em methods\/} of manipulation with these data
are encapsulated in common structure: {\em an object}. Basic elements of FP
are functions, but the same structure like {\sf (add~1~x)}
can be treated either as a data (the {\em list\/} with three elements: $add$,
$1$, $x$) or, as the body of a {\em function\/} (the addition of $1$ to $x$).

 From point of view of OOP the functions in FP can be considered as special
type of {\em functional objects}.
It was mentioned above that such {\em ``lazy functions''\/} are essential for
symbolic manipulation. It can be useful to add this conception to OOP\@.
Internal realization of such function is some {\em FP Object\/}:
{\obeylines
\ $FP\!\_function$ = \underline{\bf Object}($FP\!\_Object$)
\quad Arg : Arguments;
\quad       \underline{\bf function} Evaluate : Result;
\ \underline{\bf end}; \{ $\underline{FP\!\_function}$ F(Arguments) : Result \}
}

\subsection{Type checking}

We should also to save the clear principles of {\em types\/} in OOP\@.
For example, the expression $y:=3*(x+1)$ must be valid. It means that
type of $3$ ({\bf integer}) is compatible with functional object
{\sf (Arg:(1,x),+)} i.e.\ any {\em functional object\/} has {\em type\/}
compatible with {\em type of result\/} of the function. The {\em variable\/}
$x$ in this expression does not have an assigned value. Such kind of objects
are useful for symbolic manipulation and it has used in FP or {\em logical
programing}, but it lack of in usual imperative OOP languages. We should
distinguish such variable and variable with value, but both types must be
compatible.

For using of the FP principles in \ppp\ are introduced some
extensions of {\em types\/}. Usually if we have some variable
of type {\sf T} we can assign to it only value of the same or descendant
type. In \ppp\ there are three variant for every type {\sf T}:
\underline{\bf value} of {\sf T}, \underline{\bf variable} of {\sf T}, and
\underline{\bf functional object} of {\sf T}. For example:

\medskip

{\obeylines
 \underline{\bf var}
 \quad $a, b, c, d : $ \underline{\bf integer}; %
 \{ $a, b, c, d : $ \underline{\bf integer} \underline{\bf variable} \}
 \quad $a := 1$;\ \{ $a : $ \underline{\bf integer} \underline{\bf value} \}
 \quad $b := c + d$;%
 \ \{ $b : $ \underline{\bf integer} \underline{\bf functional object} \}
}

\subsection{Algebraic example}
\noindent
After such extension we can continue an analogy with language of mathematics.
We can describe property of {\em distributivity\/} for object {\em Algebra}.

\medskip

{\obeylines
\ \underline{\bf function} Algebra.\underline{\bf infix}$*$ ($A, B$ : Algebra) : Algebra;
\ \underline{\bf par}
\quad $C, D, E, F$ : Algebra;
\ \underline{\bf begin}
\quad \underline{\bf if} $A = C + D$ \underline{\bf then}%
\quad \{ {\em i.e.} $A$ is record $(+,C,D)$ \}
\ {\bf Return} $ := C * B + D * B$
\quad \underline{\bf else}
\qquad \underline{\bf if} $B = E + F$ \underline{\bf then}%
\ {\bf Return} $ := A * E + A * F$
\qquad \underline{\bf else} {\bf Return} $ := $ {\bf fail}
\ \underline{\bf end};
}

\medskip

The {\bf fail} is value compatible with all types. In OOP it is
useful for {\em inheritance}. For example, for object {\em Complex\/} is
possible to write:

\medskip
\medskip

{\obeylines
\ \underline{\bf function} Complex.\underline{\bf infix}$*$ (A, B : Complex) : Complex;
\ \underline{\bf begin}
\quad{\bf Return} $ := $ Algebra.$(A * B);$ \{ Inherited method for algebraic expressions \}
\quad \underline{\bf if} {\bf Return} $ = $ {\bf fail} \{ No -- Directly perform the multiplication \}
\quad \underline{\bf then} {\bf Return} $ := (A.Re * B.Re - A.Im * B.Im, A.Re * B.Im + A.Im * B.Re)$
\ \underline{\bf end};
}

\medskip

After such definition we can freely use the same operation for manipulation
with algebraic expression and with usual complex numbers. The calling of
inherited part of method is managed with algebraic transformations like
distributivity \mbox{($(i + x) * i \to i * i + i * x $)} and another part perform
usual complex multiplication \mbox{($ i * i + i * x \to -1 + i * x $)}.


\end{document}